\RequirePackage{fix-cm}
\documentclass[smallextended]{svjour3}       % onecolumn (second format)
\smartqed  % flush right qed marks, e.g. at end of proof
\usepackage{graphicx}
\usepackage{cite}
\begin{document}

\title{Unidimensional continuous-variable measurement-device-independent quantum key distribution%\thanks{Grants or other notes
}


\author{Dongyun Bai \and Peng Huang\thanks{Corresponding author: huang.peng@sjtu.edu.cn} \and Yiqun Zhu\thanks{Corresponding author: zhuyiq@sdju.edu.cn} \and Hongxin Ma \and Tailong Xiao \and Tao Wang \and Guihua Zeng %etc.
}


\institute{
	Dongyun Bai 
	\at State Key Laboratory of Advanced Optical Communication Systems and Networks, School of Physics and Astronomy, Shanghai Jiao Tong University, Shanghai 200240, China
	\and
	Dongyun Bai \and Peng Huang \and Hongxin Ma \and Tailong Xiao \and Tao Wang \and Guihua Zeng 
	\at State Key Laboratory of Advanced Optical Communication Systems and Networks, Center of Quantum Sensing and Information Processing, Shanghai Jiao Tong University, Shanghai 200240, China
	\and
	Yiqun Zhu
	\at School of Electronic Information, Shanghai Dianji University, Shanghai 201306, China
}

\date{\today}

\maketitle

\begin{abstract}
Continuous-variable (CV) measurement-device-independent (MDI) quantum key distribution (QKD) is immune to imperfect detection devices, which can eliminate all kinds of attacks on practical detectors. Here we first propose a CV-MDI QKD scheme using unidimensional modulation (UD) in general phase-sensitive channels. The UD CV-MDI QKD protocol is implemented with the Gaussian modulation of a single quadrature of the coherent states prepared by two legitimate senders, aiming to simplify the implementation compared with the standard, symmetrically Gaussian-modulated CV-MDI QKD protocol. Our scheme reduces the complexity of the system since it ignores the requirement in one of the quadrature modulations as well as the corresponding parameter estimations. The security of our proposed scheme is analyzed against collective attacks, and the finite-size analysis under realistic conditions is taken into account. UD CV-MDI QKD shows a comparable performance to that of its symmetrical counterpart, which will facilitate the simplification and practical implementation of the CV-MDI QKD protocols.      

\end{abstract}

\section{Introduction}\label{intro}

Quantum key distribution (QKD) \cite{bennett1984update, ekert1991quantum, gisin2002quantum, scarani2009the, braunstein2005quantum, Liao2017Satellite} is one of the most mature applications in quantum information processing and quantum cryptography technology, which guarantees the unconditional secure key distribution between two remote partners, named Alice and Bob, even with the existence of a potential eavesdropper named Eve. The unconditional security is provided by the basic physical principles of quantum mechanics \cite{bang2006quantum}. Continuous-variable (CV) QKD protocols \cite{ralph1999continuous, grosshans2002continuous, grosshans2003quantum, bai2017performance, liu2018integrating}, as counterparts of the discreet-variable (DV) protocols \cite{lo2005decoy, xuan2009a, lo2014secure} where key information is encoded on the properties of single photons, have emerged advantages in high secret key rates and superior compatibility with practical optical systems. CV-QKD protocols can be implemented with the Gaussian modulation of the field quadratures of coherent states or squeezed states of light \cite{gottesman2003secure, garcia2009continuous}. In the last two decades, researches on CV-QKD have gradually matured \cite{weedbrook2012gaussian}. In theory, Gaussian-modulated coherent-state (GMCS) CV-QKD protocols have been proved to be secure under collective attacks \cite{garcia2006unconditional, navascues2006optimality} and coherent attacks \cite{furrer2012continuous, leverrier2013security}, even with finite-size regime \cite{leverrier2010finite, jouguet2012analysis} and composable security \cite{leverrier2015composable} taken into full analysis. Numerous experimental realizations in the laboratory \cite{lodewyck2007quantum, jouguet2013experimental, qi2015generating, wang2018high} and several field tests \cite{fossier2009field, jouguet2012field, huang2016field} have been achieved, which show the feasibility and practicability of CV-QKD protocols. A recent experiment of all-fiber GMCS CV-QKD has achieved the secure transmission distance beyond 100 km under laboratory conditions, which will contribute to the realization of metropolitan quantum networks with conventional telecom technologies \cite{huang2016long-distance}.

Theoretically, the CV-QKD protocols with Gaussian modulation have been proved to be unconditionally secure under some ideal assumptions. However, in practical implementations, imperfect devices especially practical detectors may lead to some potential loopholes \cite{gerhardt2011full}, which will further hinder the development of CV-QKD protocols. More recently, quantum attack strategies against practical detection such as local oscillator (LO) fluctuation attack \cite{Ma2013Local}, LO calibration attack \cite{Jouguet2013Preventing}, saturation attack \cite{qin2013saturation} and homodyne-detector-blinding attack \cite{QHHDBACVQKD2018}, will seriously deteriorate the actual performance of the practical quantum communication systems. One natural solution is to find a counterpart to every specific loophole, while it cannot prevent an unknown attack effectively and will greatly increase the complexity of physical implementation. To effectively fill the gap between the ideal assumptions and practical implementations, measurement-device-independent (MDI) QKD protocols were first proposed by two groups independently \cite{Braunstein2012Side, lo2012measurement-device-independent}, which are immune to all side-channel attacks against detectors. Inspired by the CV entanglement swapping, the MDI framework was extended to CV systems later \cite{pirandola2015high-rate, Ma2013Gaussian, li2014continuous-variable}. CV-MDI QKD was theoretically introduced in detail with free-space experimental proofs in Ref \cite{pirandola2015high-rate} . In most CV-MDI QKD protocols, both Alice and Bob are legitimate senders, and they perform symmetrical Gaussian modulations on amplitude and phase quadratures of coherent states. Then they send their quantum states to an untrusted third party named Charlie,  who performs Bell-state measurement (BSM) and then communicates the results to establish a secure key. Since the detection is carried out by the untrusted third party, the quantum attacks related to detectors will naturally be removed, which shows the high practical security of CV-MDI QKD protocols. Till now, several tremendous results \cite{ma2018continuous, zhao2018continuous, wang2018self, yin2019phase, ma2019long, bai2019passive} have been obtained under the theoretical framework of CV-MDI QKD, with finite-size analysis \cite{papanastasiou2017finite, zhang2017finite} and composable security analysis \cite{lupo2018continuous} fully accomplished.

A further simplified unidimensional modulation (UD) CV-QKD protocol has been proposed to reduce the system complexity and the cost of the apparatus \cite{usenko2015unidimensional}, which thereby facilitate the commercialization of practical CVQKD schemes. Compared to the conventional symmetrical GMCS CV-QKD protocols, the asymmetrical UDCV-QKD protocols only requires the sender to use one simple modulator to perform a single-quadrature modulation instead of two modulators, which would even avoid to create a \textit{hole} in the center of the Gaussian probability distribution \cite{usenko2015unidimensional}. Moreover, the security analysis \cite{wang2017finite, liao2018composable, wang2018security} and several experimental realizations \cite{wang2017experimental} were carried out to validate the feasibility of UDCV-QKD protocols. 

So far, in all presented CV-MDI QKD protocols \cite{pirandola2015high-rate, li2014continuous-variable}, two senders both propose a symmetrical modulation by using amplitude and phase modulators, which causes the CV-MDI QKD protocols relatively complex. In order to reduce the complexity of CV-MDI QKD protocols, in this paper we extend the idea of UD to CV-MDI QKD framework, and we firstly propose a CV-MDI QKD protocol based on unidimensional modulation. In this renewed scheme, both Alice and Bob use one modulator to finish the single-quadrature modulation, then they send their prepared quantum states to Charlie for BSM. We analyze the security in a general phase-sensitive Gaussian channel under optimal collective attacks \cite{usenko2015unidimensional}. Under the physicality constraints and rational parameters related to unmodulated quadrature, we obtain the secret key rates in our UD CV-MDI QKD protocol. We also take the finite-size effects into our security analysis to obtain a tight bound under practical conditions.

The paper is structured as follows. In Sect. \ref{sec2}, we first review the original UD CV-QKD structure and the illustration of symmetrical modulated CV-MDI QKD protocols. In Sect. \ref{SKCPA}, we derive the secret key rate of the UD CV-MDI QKD protocol in asymptotic case, in  comparisons with the conventional, symmetrical modulated CV-MDI QKD protocols. The finite-size analysis is fully taken into account in Sect. \ref{FSAUDCVMDI}. Finally the conclusions and discussions are drawn in Sect. \ref{Con}.                             

\section{CV-MDI QKD protocol with unidimensional modulation}
\label{sec2}

In this section, we first review the UDCV-QKD protocol and the original CV-MDI QKD protocol with symmetrical modulation. Then we introduce our proposed UD CV-MDI QKD protocol with the equivalent entanglement-based (EB) scheme presented in detail, which is more convenient and reasonable to perform the security analysis. 

\subsection{UDCV-QKD protocol and original CV-MDI QKD protocol}
\label{sec2.1}

\begin{figure}[!h]\center
	\resizebox{12cm}{!}{
	\includegraphics{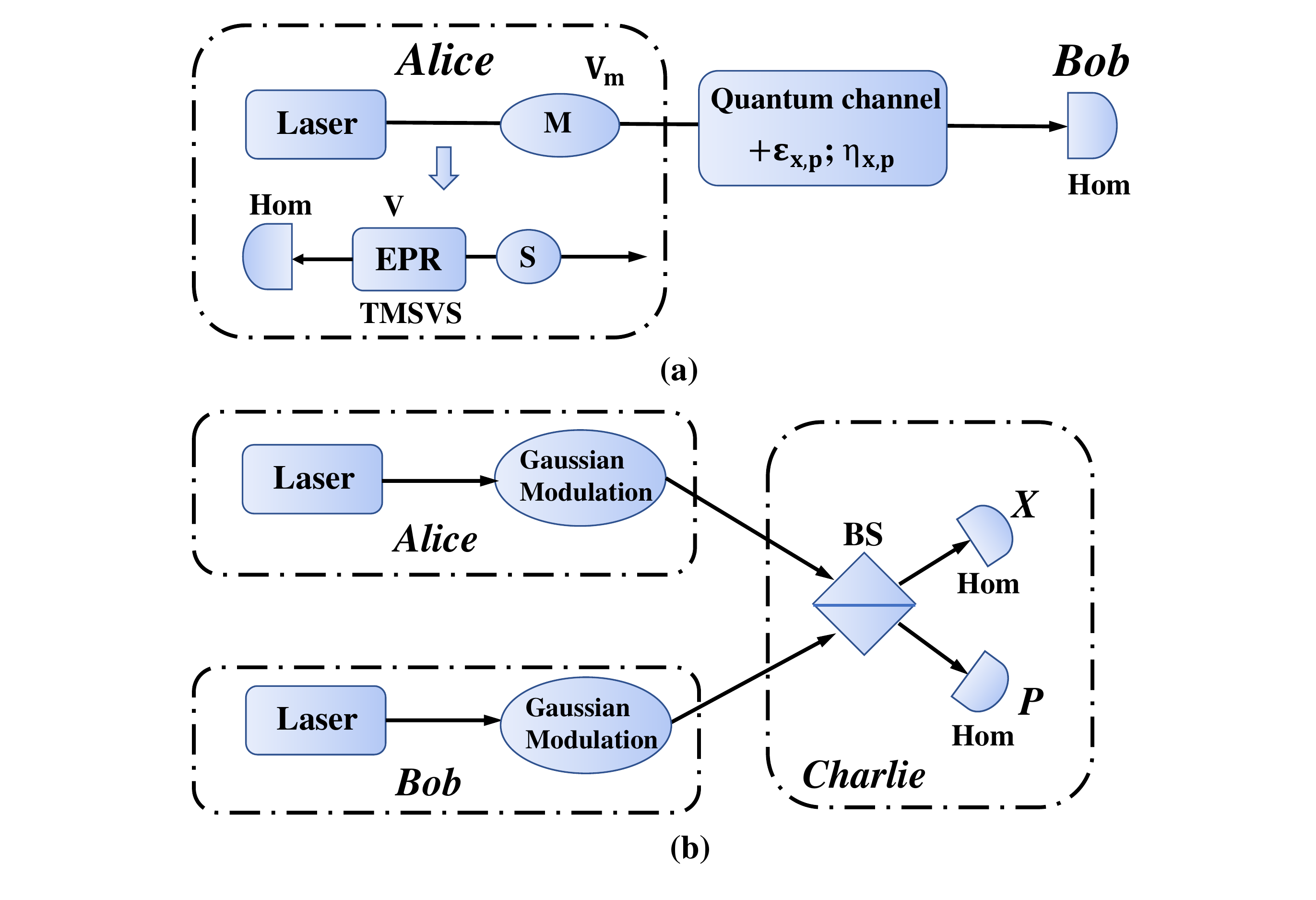}}
	\caption{(a) The prepare-and-measure (PM) model and equivalent EB model of general UDCV-QKD protocol. (b) PM model of original CV-MDI QKD protocol. M: one single modulator; S: squeezer; EPR: Einstein-Podolsky-Rosen state; TMSVS: two-mode squeezed vacuum state; Hom: homodyne detection; BS: beam splitter.}
	\label{fig:2.1}       % Give a unique label
\end{figure}

The schematic of UDCV-QKD protocol is displayed in Fig. \ref{fig:2.1}(a). In prepare-and-measure (PM) model, the trusted sender Alice modulates one quadrature (amplitude quadrature $\hat{x}$ or phase quadrature $\hat{p}$) of the coherent states (generated from a laser source) with modulation variance $V_{m}$ by one single modulator M and then she distributes the quantum states to the remote trusted party Bob. Bob implements homodyne detection to detect the modulated quadrature. Alice and Bob use reverse reconciliation to extract secret keys by data post-processing method. Without loss of generality, in the rest of our paper, we further assume that the senders modulate the amplitude quadrature $\hat{x}$. The quantum channel is characterized as a phase-sensitive channel, with transmittance $\eta_{x,p}$ and excess noise $\epsilon_{x,p}$ in $\hat{x}$ and $\hat{p}$ quadratures respectively. It should be noted that the receiver needs to measure the other unmodulated quadrature $\hat{p}$ sometimes to acquire the necessary properties of the channel in $\hat{p}$ quadrature. In EB model, Alice measures one half mode of a two-mode squeezed vacuum state (TMSVS) with variance V by homodyne detection, while the other half mode is squeezed by the squeezer S and then it's projected into a coherent state and sent to the quantum channel \cite{usenko2015unidimensional, wang2017experimental} to extract a secret key.  

Figure \ref{fig:2.1}(b) shows the PM version of the conventional CV-MDI QKD protocol \cite{pirandola2015high-rate}. The main procedures can be briefly described in the following ways: (1) Both Alice and Bob prepare the quantum states independently with symmetrical Gaussian modulation on amplitude and phase quadratures. Then the prepared quantum states are sent to the untrusted third party Charlie through two independent quantum links. (2) Charlie performs BSM on the incoming modes by interfering them on a balanced beam splitter (BS). The two output modes from BS are measured by two homodyne detectors, with the results of $\hat{X}$ and $\hat{P}$ announced publicly. (3) Alice and Bob use the measurement results to modify their data and then to establish a string of raw keys. (4) Alice and Bob implement parameter estimation, information reconciliation and privacy amplification to finally obtain a string of secret keys.             

\subsection{CV-MDI QKD with unidimensional modulation}
\label{sec2.2}

\begin{figure}[!h]\center
	\resizebox{12cm}{!}{
		\includegraphics{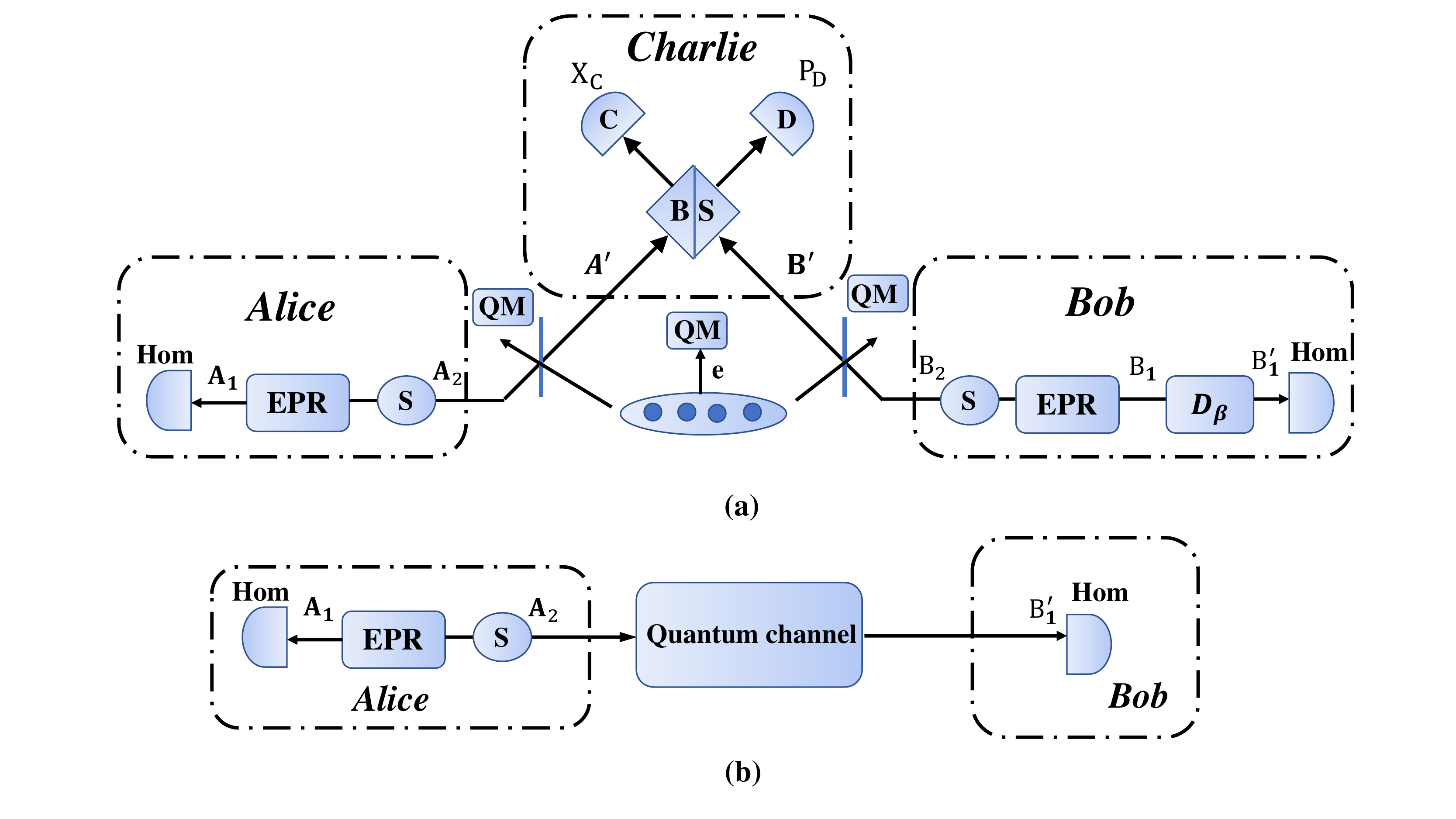}}
	\caption{(a) The equivalent EB version of CV-MDI QKD with unidimensional modulation. (b) Simplified equivalent one-way protocol of CV-MDI QKD protocol with unidimensional modulation. M: one single modulator; S: squeezer; EPR: Einstein-Podolsky-Rosen state; Hom: homodyne detection; BS: beam splitter; QM: quantum memory. $D_{\beta}$ is displacement operation.}
	\label{fig:2.2}       % Give a unique label
\end{figure}

In this part, we come to the implementation of CV-MDI QKD protocol with unidimensional modulation based on its equivalent EB version as illustrated in Fig. \ref{fig:2.2}(a), for the convenience of security analysis. First we consider the state preparation at Alice's side, where a TMSVS with variance V is prepared (EPR state). One mode of the EPR state is squeezed by a squeezing operation $S$ with a squeezing parameter $\mathrm{-log\sqrt{V}}$, while the other half mode $A_{1}$ is measured by Alice using homodyne detection. Then mode $A_{2}$ is conditionally prepared in coherent states with modulation variance $V_{m} = V^{2} - 1$ and sent to Charlie. Similarly, Bob preforms the same unidimensional modulation and sends mode $B_{2}$ to Charlie. Mode $A'$ and mode $B'$ are interfered at a balanced BS at Charlie's side. Under the previous and agreed  assumption that the single modulated quadrature is amplitude quadrature, Charlie announces the $\hat{x}$ quadrature of $C$ publicly. After receiving the measurement results of Charlie, Bob displaces mode $B_{1}$ through displacement operation $D_{\beta}$ and gets $\hat{\rho}_{B'_{1}} = D_{\beta}\hat{\rho}_{B_{1}}D^{\dagger}_{\beta}$, where $\hat{\rho}$ represents the density matrix operator while $\beta$ is related to the gain of displacement of Charlie's measurement results, and Alice keeps her measured data unchanged. Finally, Alice and Bob use an authenticated channel for parameter estimation, reverse reconciliation and privacy amplification to obtain a string of secure keys.

Here are two points needed to emphasize. One point is that after Charlie's measurements and Bob's displacement, mode $A_{1}$ and mode $B'_{1}$ can be treated entangled and their data is then correlated \cite{li2014continuous-variable}. The other point is that Alice and Bob ought to sometimes switch $\hat{X}$ and $\hat{P}$ basis and modulate the phase quadrature $\hat{p}$, then Charlie needs to reveal the interference results of $P_{D}$ of $D$ sometimes for both senders to gather essential channel properties in quadrature $\hat{p}$ \cite{usenko2015unidimensional, liao2018composable}.           

\section{Secret key calculation with performance analysis under asymptotic case}
\label{SKCPA}

\subsection{Secret key calculation}
\label{SKC}

In this section, we mainly carry out the secret key rate analysis under Gaussian collective attacks for they are proved to be optimal in asymptotic case. We derive secret key rate based on the EB scheme in Fig. \ref{fig:2.2}(a). When the EPR state at Bob's side and the displacement $D_{\beta}$ are accessible to Eve, the EB scheme in Fig. \ref{fig:2.2}(a) can be conveniently equivalent to a common one-way UDCV-QKD protocol \cite{li2014continuous-variable} shown in Fig. \ref{fig:2.2}(b). For the equivalent one-way model requires more constraints on Eve, it is obvious the secret key rate in Fig. \ref{fig:2.2}(b) is a lower bound of that derived from Fig. \ref{fig:2.2}(a). To facilitate the calculation process with covariance matrix, we use the model in Fig. \ref{fig:2.2}(b) to obtain our $K_{UD}$ under collective attacks.

The CV-MDI QKD protocol has two quantum channels and there exists two main eavesdropping strategies: one-mode attack and two-mode attack. In practice, it's challenging for Eve to obtain quantum correlations from both channels due to technical constraints. In our work, we restrict our channels to two independent Gaussian Markovian memoryless channels, where Eve can fully implement one-mode attack. However, we should point out that Eve's attack assumed here is not as optimal as two-mode attack in Ref \cite{pirandola2015high-rate}.

Generally, the lower bound of the secret key rate in Fig. \ref{fig:2.2}(b) under optimal collective attack can be given as
\begin{eqnarray}
\label{eq1}
K_{UD} &=& \beta I_{A_{1}B'_{1}} - \chi_{E} \nonumber \\
&=& \beta I_{A_{1}B'_{1}} - (S(E) - S(E|X_{B'_{1}})). 
\end{eqnarray}        
where $\beta$ is the reconciliation efficiency, $I_{A_{1}B'_{1}}$ is the Shannon mutual information between Alice and Bob with $\chi_{E}$ the Holevo bound between Bob and Eve, $S$ represents the Von Neumann entropy. Since Eve could purify the whole system after Bob performs homodyne detection, thus the mutual information between Bob and Eve can expressed as $\chi_{E} = S(A_{1}B'_{1}) - S(A_{1}|X_{B'_{1}})$.   
%\paragraph{Paragraph headings} Use paragraph headings as needed.

In our UD CV-MDI QKD protocol, we focus on the modulation of $\hat{x}$ quadrature, which results in asymmetrical covariance matrix compared with its symmetrical Gaussian modulation CV-QKD protocol. We assume that Alice and Bob use the same modulation variance $V_{m}$ and the transmittance and excess noise in Alice's (Bob's) channel are $\eta_{A}(\eta_{B})$ and $\epsilon_{A}(\epsilon_{B})$. In the EB scheme in Fig. \ref{fig:2.2}(b), Alice measures one mode of EPR state of variance V, while the other half mode is squeezed with the squeezing parameter $\mathrm{-log\sqrt{V}}$, which results the covariance matrix as:
\begin{equation}
\label{array1}
    \gamma_{A_{1}A_{2}} = \left( {\begin{array}{*{20}{c}}
	V& 0 & {\sqrt{V(V^{2}-1)}} & 0 \\ 
	0& V & 0 & -\sqrt{\frac{V^{2}-1}{V}} \\ 
	{\sqrt{V(V^{2}-1)}}& 0 & {V^{2}} & 0 \\ 
	0& {-\sqrt{\frac{V^{2}-1}{V}}} & 0 & 1
	\end{array} } \right).
\end{equation}
Then the EB scheme is equivalent to modulate the quadrature $\hat{x}$ with modulation variance $V_{m} = V^{2} -1$. After the prepared states are sent to Bob through the quantum channel with transmittance $\eta_{x,p}$ and excess noise $\epsilon_{x,p}$, the covariance matrix $\gamma_{A_{1}A_{2}}$ is transformed into covariance matrix $\gamma_{A_{1}B'_{1}}$ in the following form assuming perfect homodyne detection:
\begin{eqnarray}
\label{array2}
 \left( \begin{small}{\begin{array}{*{10}{c}}
	V& 0 & \sqrt{T_{A,x}}{\sqrt{V(V^{2}-1)}} & 0 \\ 
	0& V & 0 & {-\sqrt{T_{A,p}}\sqrt{\frac{V^{2}-1}{V}}} \\ 
	{\sqrt{T_{A,x}}\sqrt{V(V^{2}-1)}}& 0 & {T_{A,x}(V^{2}-1+\epsilon'_{A,x})+1} & 0 \\ 
	0& {-\sqrt{T_{A,p}}\sqrt{\frac{V^{2}-1}{V}}} & 0 & {1+T_{A,p}\epsilon'_{A,p}}
	\end{array} } \end{small} \right)
\end{eqnarray}
where $V = \sqrt{V_{m}+1}$, $T_{A,x}$ and $\epsilon'_{x}$ correspond to Alice's channel estimated parameters in $\hat{x}$ quadrature, and they are given as:
\begin{eqnarray}
\label{array3}
T_{A,x} = \frac{\eta_{A,x}}{2}g^{2},
\end{eqnarray}
while
\begin{eqnarray}
\label{array4}
\epsilon'_{A,x} = 1 &+& \frac{\eta_{B,x}}{\eta_{A,x}}(\chi_{B,x} - 1) + \chi_{A,x}  \nonumber \\ &+& \frac{1}{\eta_{A,x}}(\frac{\sqrt{2V_{m}}}{g}-\sqrt{\eta_{B}(V_{m}+2)})^{2},
\end{eqnarray}
with $\chi_{A,x} = \frac{1-\eta_{A,x}}{\eta_{A,x}}+\epsilon_{A,x}$, $\chi_{B,x} = \frac{1-\eta_{B,x}}{\eta_{B,x}}+\epsilon_{B,x}$, g is the gain of the displacement $D_{\beta}$ in Bob's side. To minimize the excess noise $\epsilon'_{A,x}$, we choose $g^{2} = \frac{2V_{m}}{\eta_{B,x}(V_{m}+2)}$ and derive:
\begin{eqnarray}
\label{array5}
\epsilon'_{A,x} = \epsilon_{A,x} + \frac{2}{\eta_{A,x}} +\frac{\eta_{B,x}}{\eta_{A,x}}(\epsilon_{B,x} - 2).
\end{eqnarray}
 
While $T_{A,p}$ and $\epsilon'_{A,p}$ is correlated with the $\hat{p}$ quadrature. Theoretically, before we derive the explicit  expressions of the parameters, we have to consider the relationship of the two unknown parameters $\eta_{A,p}$ and $\epsilon_{A,p}$ in $\hat{p}$ quadrature. Bounded by the Heisenberg uncertainty principle to meet the requirement of physicality, the two unknown parameters should satisfy the parabolic equation constraint \cite{wang2018security}:
\begin{equation}
\label{array6}
(\sqrt{\frac{\eta_{A,x}}{(1+\eta_{A,x}\epsilon_{A,x})^{2}}}-\sqrt{\eta_{A,p}})^{2} \leq (1-\frac{\eta_{A,x}}{1+\eta_{A,x}\epsilon_{A,x}})(1+\eta_{A,p}\epsilon_{A,p}-\frac{1}{\eta_{A,x}\epsilon_{A,x}}). 
\end{equation}   

\begin{figure}[!h]\center
	\resizebox{12cm}{!}{
		\includegraphics{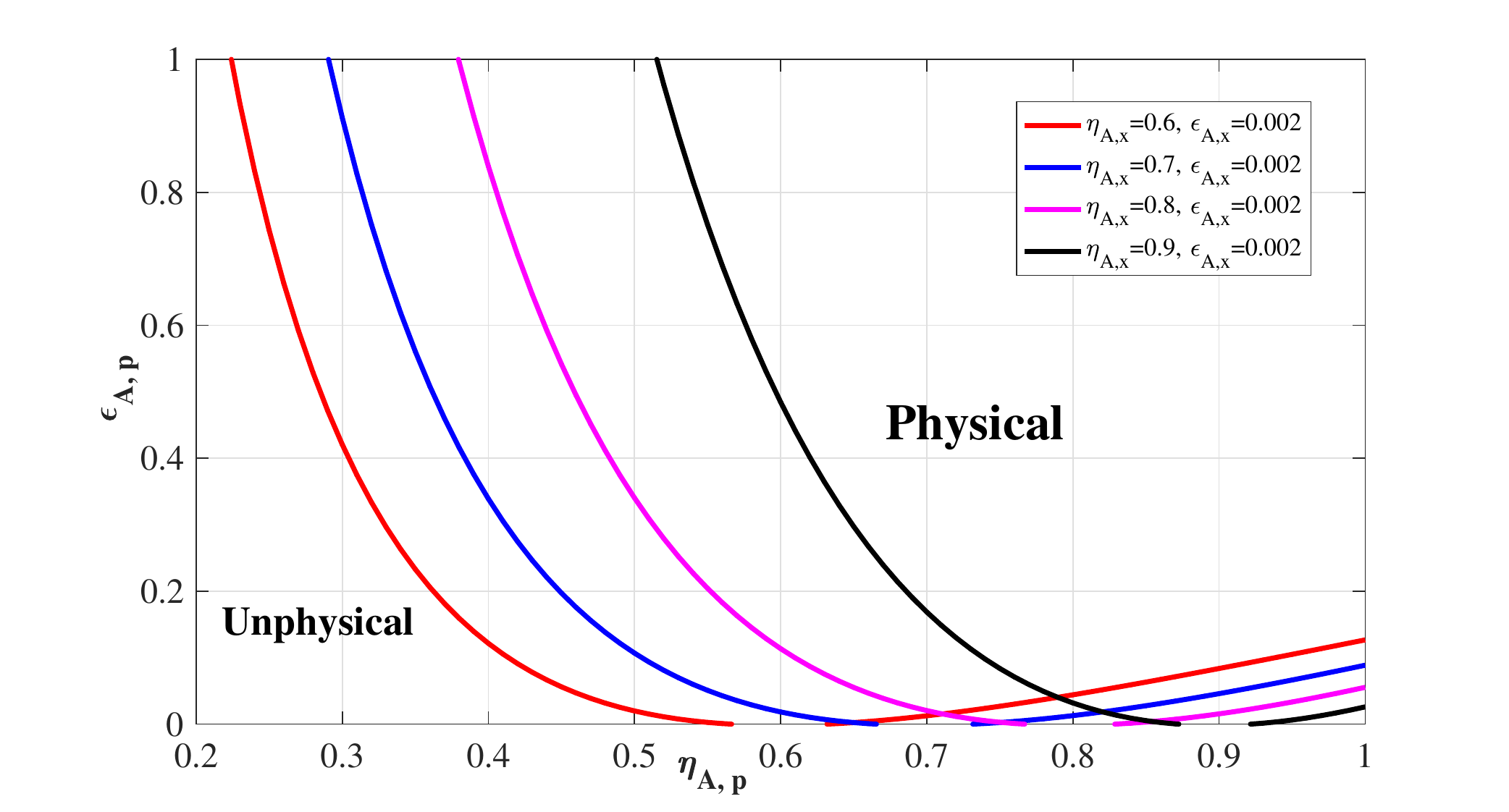}}
	\caption{Regions bounded by physicality of the varied $\eta_{A,x}$ and $\epsilon_{A,x}$. The values of these parameters can be accessible in practice.}
	\label{fig:2.3}       % Give a unique label
\end{figure}

In Fig. \ref{fig:2.3} we explore the regions bounded by physicality with a series of parameters $\eta_{A,x}$ and $\epsilon_{A,x}$. The region is divided into three regions by every curve and on the top part separated by every individual curve, it belongs to the physical region, which means the two unknown parameters can be physically set simultaneously, otherwise the other parts will violate Heisenberg uncertainty. In typical communication channels, one always expect the values of channel loss and excess noise in both $\hat{X}$ and $\hat{P}$ quadratures are symmetric, and therefore we assume $\eta_{A,p} = \eta_{A,x}$ ($\eta_{B,p} = \eta_{B,x}$) and $\epsilon_{A,p} = \epsilon_{A,x}$ ($\epsilon_{B,p} = \epsilon_{B,x}$) in the rest of our paper to carry out secret key calculation \cite{usenko2015unidimensional, wang2017finite}. From Fig. \ref{fig:2.3} we can prove our assumption strictly satisfy the physicality to perform unidimensional modulation and now we start to calculate the secret key rate.        

The Shannon mutual information between Alice and Bob $I_{A_{1}B'_{1}}$ can be denoted as:
\begin{eqnarray}
\label{array7}
I_{A_{1}B'_{1}} = \frac{1}{2}\mathrm{log_{2}}\frac{V_{A_{1}}}{V_{A_{1}|X_{B'_{1}}}}.
\end{eqnarray}
where $V_{A_{1}}$ is the variance of mode $A_{1}$, and $V_{A_{1}|X_{B'_{1}}}$ can be derived from the matrix $\gamma_{A_{1}|X_{B'_{1}}}$, which is calculated as \cite{fossier2009improvement}:
\begin{eqnarray}
\label{array8}
\gamma_{A_{1}|X_{B'_{1}}} = \gamma_{A_{1}} - \sigma^{T}_{A_{1}B'_{1}}(X\gamma_{B'_{1}}X)^{\mathrm{MP}}\sigma_{A_{1}B'_{1}}. 
\end{eqnarray} 
where $X = diag(1,0)$ and MP represents Moore-Penrose pseudo-inverse of a matrix. $\gamma_{A_{1}}$, $\gamma_{B'_{1}}$ and $\sigma_{A_{1}B'_{1}}$ can all derived from the decomposition of $\gamma_{A_{1}B'_{1}}$.

After some algebra calculation, we can obtain
\begin{eqnarray}
\label{array9}
 I_{A_{1}B'_{1}} = \frac{1}{2}\mathrm{log_{2}}\frac{V}{V - \frac{T_{A,x}V(V^2-1)}{T_{A,x}(V^2+\epsilon'_{A,x}-1)+1}}.
\end{eqnarray}

As we have stated before, Eve can provide a purification of the whole system, so we can derive $S(E) = S(A_{1}B'_{1})$ and $S(E|X'_{B_{1}}) = S(A_{1}|X'_{B_{1}})$. $S(A_{1}B'_{1})$ can be written as a function of the symplectic eigenvalues $\lambda_{1,2}$ of $\gamma_{A_{1}B'_{1}}$, denoted as
\begin{eqnarray}
\label{array10}
S(A_{1}B'_{1}) = G(\lambda_{1}) + G(\lambda_{2}),
\end{eqnarray}   
with
\begin{eqnarray}
\label{array11}
G(x) = \frac{(x+1)}{2}\mathrm{log_{2}}\frac{(x+1)}{2} - \frac{(x-1)}{2}\mathrm{log_{2}}\frac{(x-1)}{2}.
\end{eqnarray}
Similarly, $S(A_{1}|X'_{B_{1}})$ can be denoted as $S(A_{1}|X'_{B_{1}}) = G(\lambda_{3})$, where symplectic eigenvalue $\lambda_{3}$ can be derived from the matrix $\gamma_{A_{1}|X_{B'_{1}}}$, considering the perfect homodyne detection at both Alice's and Bob's side.

Now we have derived all the parameters to calculate secret key rate of our UD CV-MDI QKD protocol under asymptotic case.

\subsection{Performance analysis}
\label{2.22}

In CV-MDI QKD protocols, there exists two different types with respect to the position of the third party Charlie. If Charlie is in the middle of Alice and Bob, we denote it as symmetric case ($L_{AC} = L_{BC}$), while if Charlie is extremely close to one party, we name it as asymmetric case ($L_{AC} \neq L_{BC}$). 

\begin{figure}[!h]\center
	\resizebox{12cm}{!}{
		\includegraphics{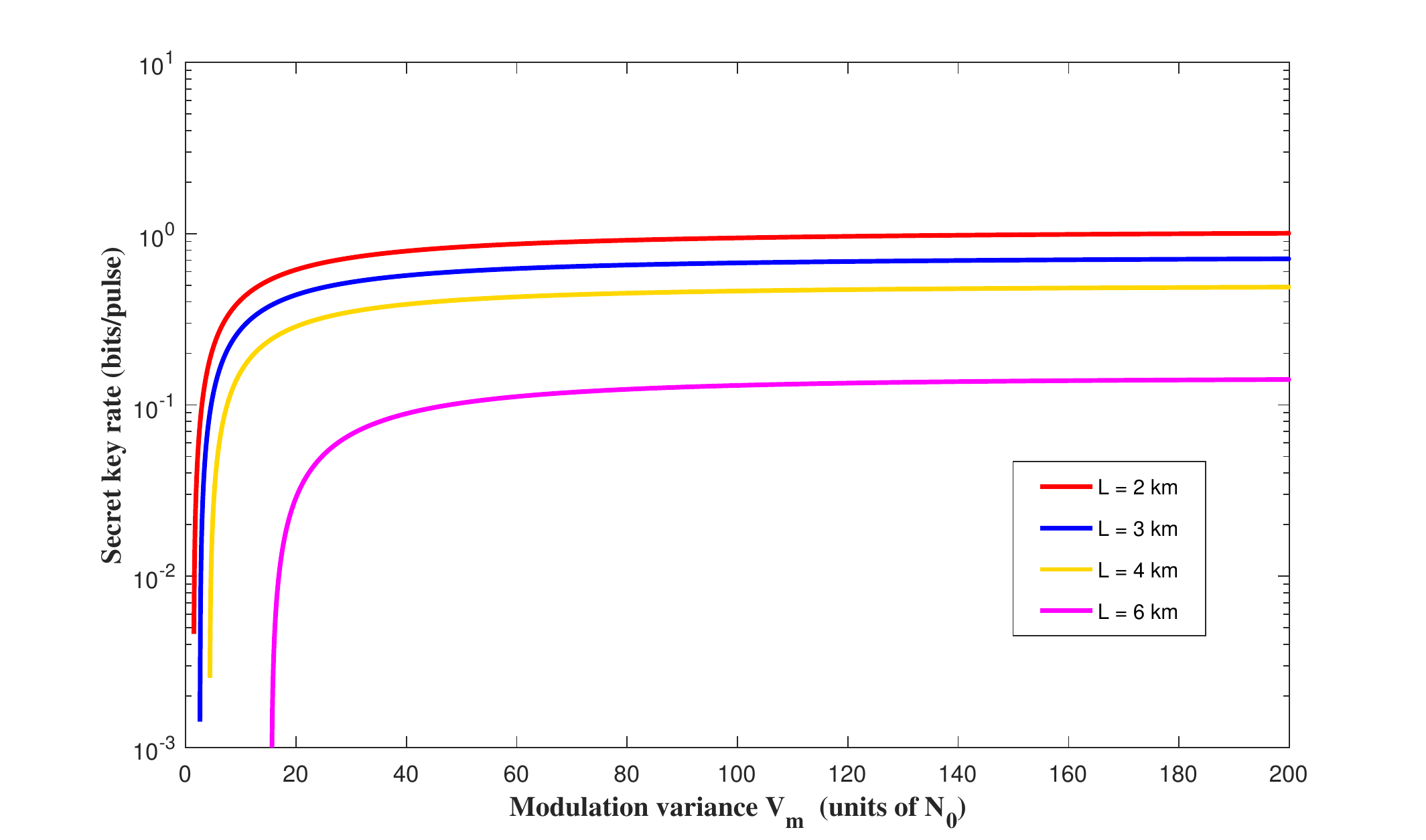}}
	\caption{Secret key rate as a function of modulation variance in symmetric case. The modulation variance $V_{m}$ is in shot noise unit $N_{0}$. The reverse reconciliation is 0.98 \cite{wang2018security}, the excess noise are $\epsilon_{A,x} = \epsilon_{B,x} = 0.002$ \cite{ma2018continuous}, the quantum channel loss is 0.2 dB/km. From top to bottom, the total transmission distance ($L = L_{AC} + L_{BC}$) is 2 km, 3 km, 4 km, 5 km. }
	\label{fig:2.4}       % Give a unique label
\end{figure}

\begin{figure}[!h]\center
	\resizebox{12cm}{!}{
		\includegraphics{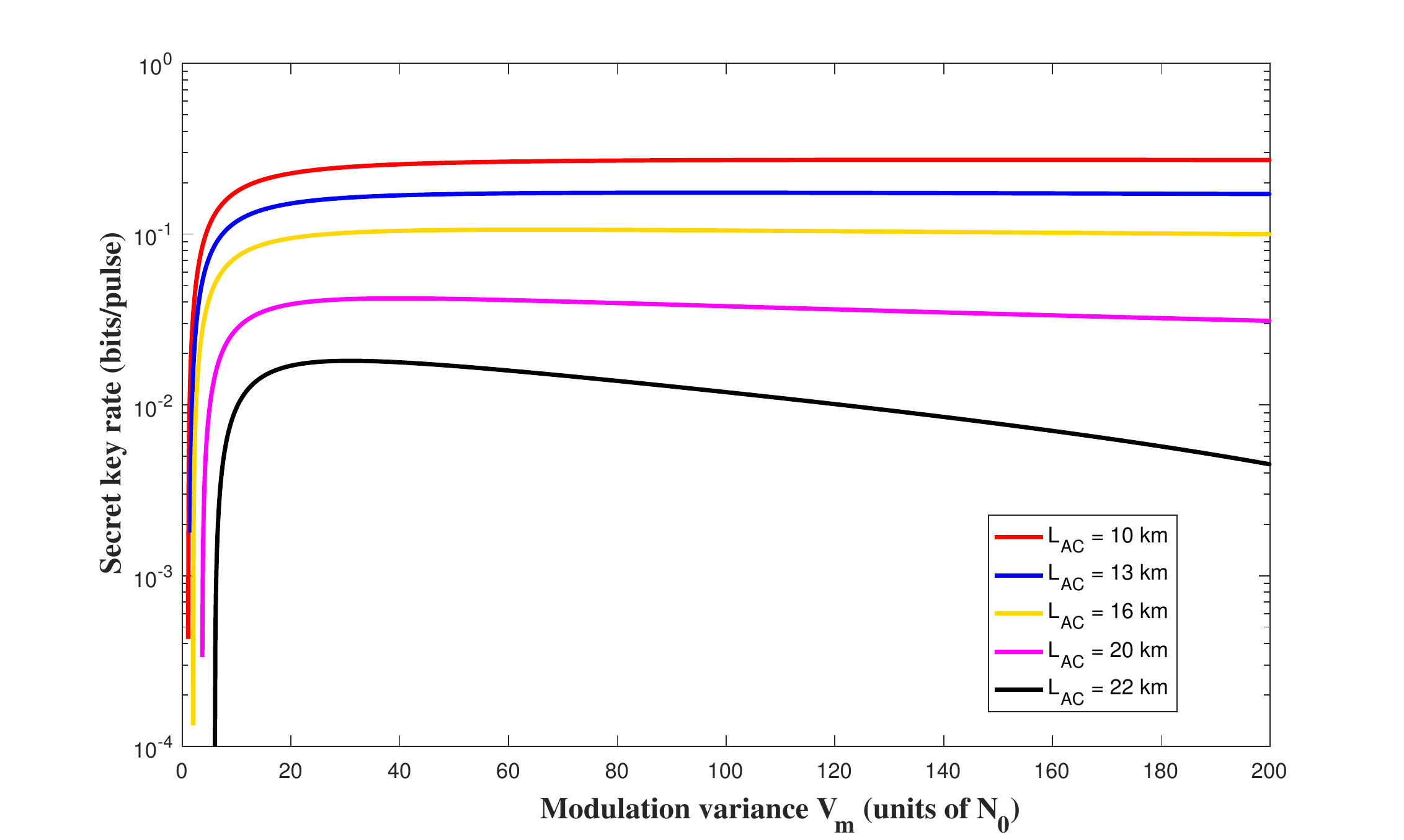}}
	\caption{Secret key rate as a function of modulation variance in asymmetric case, where Charlie is extremely close to Bob with a total efficiency $98\%$. The modulation variance $V_{m}$ is in shot noise unit $N_{0}$. The reverse reconciliation is 0.98, the excess noise are $\epsilon_{A,x} = \epsilon_{B,x} = 0.002$, the quantum channel loss is 0.2 dB/km. From top to bottom, the total transmission distance ($L_{AC}$) is 10 km, 13 km, 16 km, 20 km, 22 km. }
	\label{fig:2.5}       % Give a unique label
\end{figure}

At first, we analyze the secret key rate as a function of modulation variance $V_{m}$ since $V_{m}$ is a key parameter that will affect the performance of UD CV-MDI QKD protocol. The illustration in symmetric case is shown in Fig. \ref{fig:2.4} while the illustration in asymmetric case in shown in Fig. \ref{fig:2.5}. From the two curves, we can see that in both symmetric case and asymmetric case, the large modulation could be adopted to achieve higher secret key rate, however, when the modulation is too large, the performance is not greatly improved. So considering the practical conditions, we choose the modulation variance $V_{m} = 100$ (in shot noise unit $N_{0}$) to calculate our secret key rate, which can lead to optimal performance.  

\begin{figure}[!h]\center
	\resizebox{12cm}{!}{
		\includegraphics{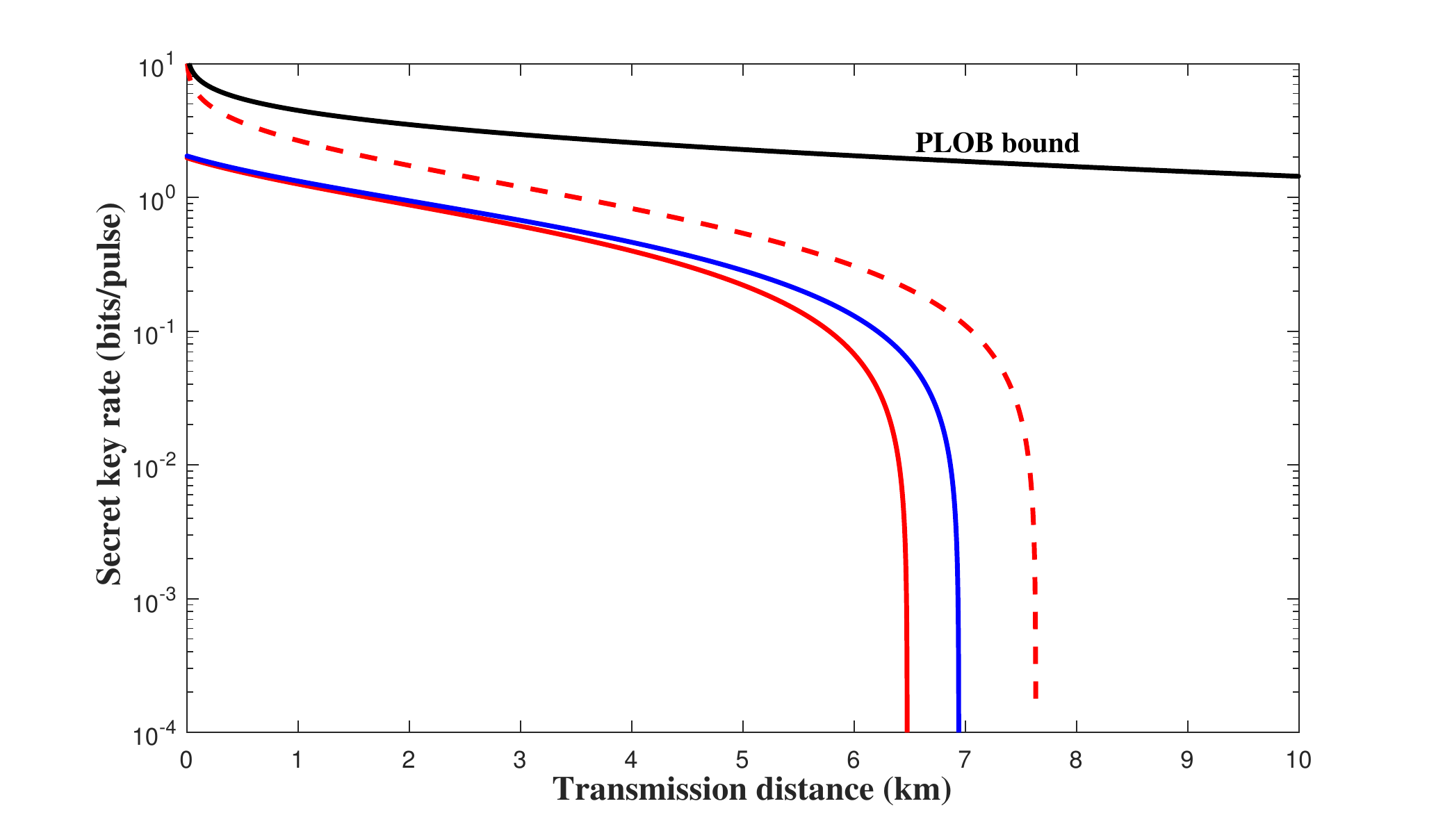}}
	\caption{Secret key rate as a function of transmission distance in symmetric case. From left to right, the red solid line represents UD CV-MDI QKD protocol with reverse reconciliation efficiency $96\%$, the blue solid line represents UD CV-MDI QKD protocol with reverse reconciliation efficiency $98\%$, while the dashed red line represents original, symmetrical Gaussian modulation CV-MDI QKD protocol with $98\%$ efficiency. The upper solid black line is the PLOB bound. The modulation variance $V_{m}$ is 100, the excess noise are $\epsilon_{A,x} = \epsilon_{B,x} = 0.002$, the quantum channel loss is 0.2 dB/km.   }
	\label{fig:2.6}       % Give a unique label
\end{figure}

The plots in Fig. \ref{fig:2.6} show secret key rate as a function of transmission distance in symmetric case, for both our UD CV-MDI QKD scheme and symmetrical Gaussian modulation CV-MDI QKD scheme. The red solid line on the left refers to the UD CV-MDI QKD protocol with $\beta = 0.96$, the blue solid line in the middle refers to the UD CV-MDI QKD protocol with $\beta = 0.98$. The dashed red line on the right represents original, symmetrical Gaussian modulation CV-MDI QKD protocol with $\beta = 0.98$, and the upper solid black line is PLOB bound, which determines the ultimate limit of repeater-less quantum communication \cite{Pirandola2017Fundamental}. We can see from Fig. \ref{fig:2.6} that in symmetric case, our proposed UD CV-MDI QKD scheme can achieve high performance with optimal modulation variance and the maximum transmission distance is satisfactory compared with the symmetrical CV-MDI QKD scheme. 

\begin{figure}[!h]\center
	\resizebox{12cm}{!}{
		\includegraphics{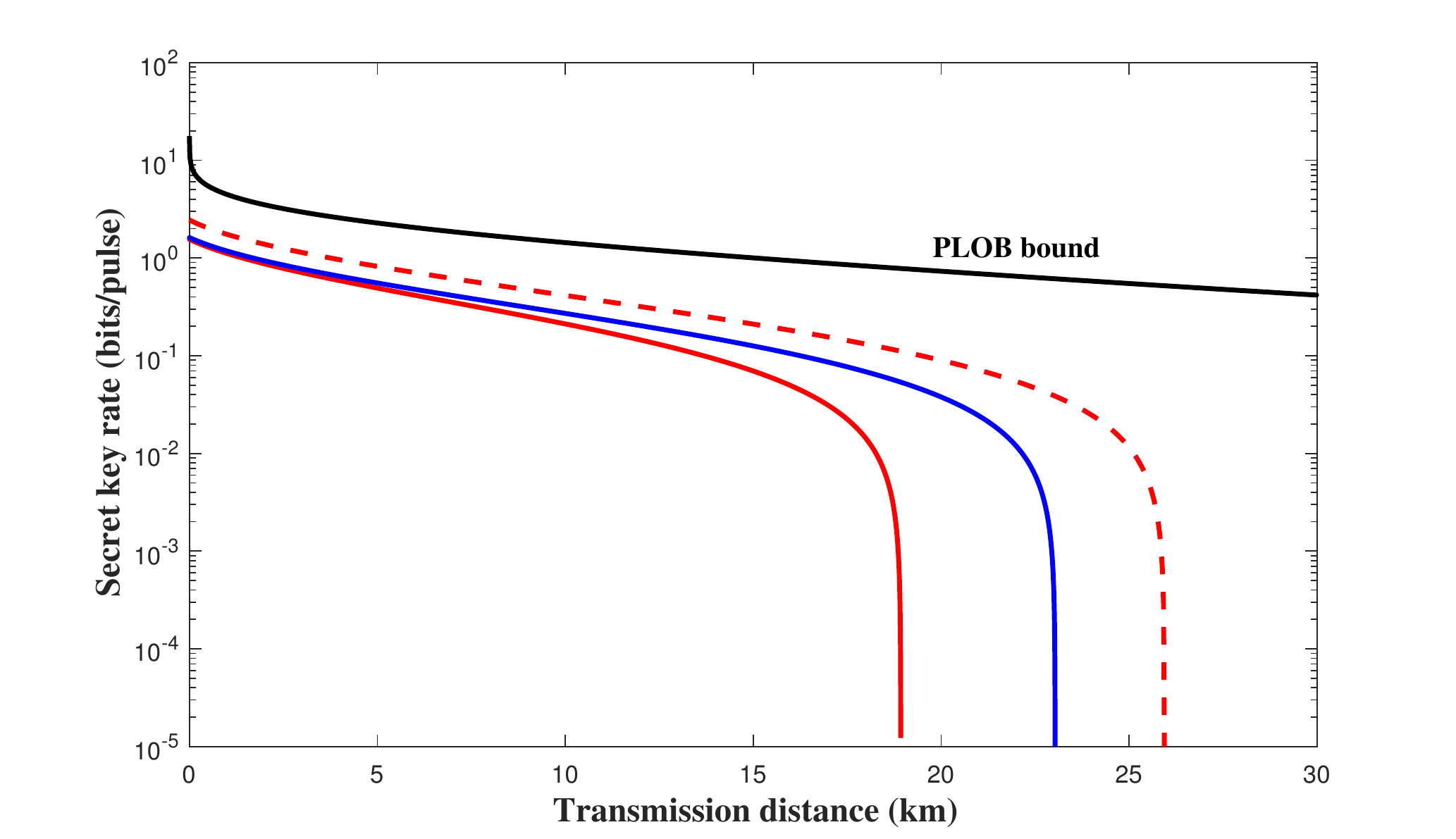}}
	\caption{Secret key rate as a function of transmission distance in asymmetric case. where Charlie is set extremely close to Bob with an overall efficiency $98\%$. From left to right, the red solid line represents UD CV-MDI QKD with reverse reconciliation efficiency $96\%$, the blue solid line represents UD CV-MDI QKD protocol with reverse reconciliation efficiency $98\%$, while the dashed red line represents original, symmetrical Gaussian modulation CV-MDI QKD protocol with $98\%$ efficiency. The upper solid black line is the PLOB bound. The modulation variance $V_{m}$ is 100, the excess noise are $\epsilon_{A,x} = \epsilon_{B,x} = 0.002$, the quantum channel loss is 0.2 dB/km.  }
	\label{fig:2.7}       % Give a unique label
\end{figure}

In the asymmetric case, the plots are drawn in Fig. \ref{fig:2.7}. The red solid line on the left refers to the UD CV-MDI QKD protocol with $\beta = 0.96$, the blued solid line in the middle refers to the UD CV-MDI QKD with $\beta = 0.98$. The dashed red line on the right refers to the original, asymmetric Gaussian modulation CV-MDI QKD with $\beta = 0.98$. The PLOB bound is plotted in the solid black line. We could get from the curve that all the plots are strictly under the PLOB bound.

From Fig. \ref{fig:2.6} and Fig. \ref{fig:2.7}, we can conclude that UD CV-MDI QKD in asymmetric case is superior to UD CV-MDI QKD in symmetric case, which has been proved in all the previous CV-MDI QKD schemes. In our proposed UD CV-MDI QKD protocols in both cases, the performance of our protocol is comparable to its corresponding original symmetric Gaussian modulation CV-MDI QKD protocol, while our protocols reduce the system complexity and simplify the implementation with more standard devices. In addition, UD CV-MDI QKD protocol is sensitive to reverse reconciliation efficiency especially in asymmetric case and it's reasonable for us to adopt more efficient reconciliation algorithms.    
%\begin{figure}[!h]\center
%% Use the relevant command to insert your figure file.
%% For example, with the graphicx package use
%  \includegraphics{example.eps}
%%\label{fig:1}       % Give a unique label
%\end{figure}
%\begin{figure*}
%% Use the relevant command to insert your figure file.
%% For example, with the graphicx package use
%  \includegraphics[width=0.75\textwidth]{example.eps}
%%\label{fig:2}       % Give a unique label
%\end{figure*}
%\begin{table}
%\begin{tabular}{lll}
%\hline\noalign{\smallskip}
%\end{tabular}
%\end{table}
\section{Finite-size analysis in UD CV-MDI QKD protocol}
\label{FSAUDCVMDI}

In the practical implementation of any CV-MDI QKD protocol, the two legitimate parties can only exchange a finite-size block of data and sacrifice part of the data for parameter estimation \cite{jouguet2012analysis, Ruppert2014Long}. To fill the gap between the protocol in asymptotic case and the protocol under practical conditions, we analyze the finite-size effects on our proposed UD CV-MDI QKD protocol in this section. As all the finite-size regime did, we mainly focus on channel transmittance and excess noise within confidence intervals. To minimize the secret key rate of our protocol, we acquire the lower transmission and higher excess noise. The secret key rate in the finite-size scenario can be expressed as:
\begin{eqnarray}
\label{array12}
K_{UD}^{f} = \frac{n}{N}[\beta I_{A_{1}B'_{1}} - S_{\epsilon_{PE}}(X_{B'_{1}}, E) - \Delta(n)].
\end{eqnarray}      
where $N$ is the total number of signals exchanged between Alice and Bob, $n$ is the number of signals used to generate secret key. The $m = N - n$ signals are used for parameter estimation. $\Delta(n)$ is the correction term related to the security of privacy amplification and has the expression
\begin{eqnarray}
\label{array13}
\Delta(n) = 7\sqrt{\frac{\mathrm{log_{2}}(2/\tilde{\epsilon})}{n}} + \frac{2}{n}\mathrm{log_{2}}(1/\epsilon_{PA}).
\end{eqnarray}
with $\epsilon_{PA}$ and $\tilde{\epsilon}$ the failure probability and the smoothing parameter. Their optimal values can be conservatively set as $\epsilon_{PA} = \tilde{\epsilon} = 10^{-10}$. $I_{A_{1}B'_{1}}$ is the mutual information of Alice and Bob, $S_{\epsilon_{PE}}(X_{B'_{1}}, E)$ is defined as the maximum entropy of Eve and Bob under certain failure probability $\epsilon_{PE}$. 

Now we come to the parameter estimation procedure and focus mainly on excess noise and transmittance. In practice, the estimation is sampled from $m$ pairs of correlated variables $(x_{i}, p_{i})_{i=1...m}$. Since the channel between Alice and Charlie, the channel between Bob and Charlie can be seen as normal linear models following Gaussian distribution. Within this model, before the BS, Alice's and Charlie's, Bob's and Charlie's data can be linked in the following relation:
\begin{eqnarray}
\label{array14}
y'_{1} &=& t'_{1}x_{1} + z_{1}, \\
y'_{2} &=& t'_{2}x_{2} + z_{2},
\end{eqnarray} 
where $t'_{1} = \sqrt{\eta_{A,x}}$, $t'_{1} = \sqrt{\eta_{B,x}}$. $z_{1}$ and $z_{2}$ follow a centered normal distribution with unknown variance $\sigma'^{2}_{1} = 1 + \eta_{A,x}\epsilon_{A,x}$ and $\sigma'^{2}_{2} = 1 + \eta_{B,x}\epsilon_{B,x}$. According to the entries of the covariance matrix, the variance of the unknown parameters before the BS can be given as:
\begin{eqnarray}
\label{array15}
\langle y'^{2}_{1} \rangle = t'^{2}_{1}V_{m} + \sigma'^{2}_{1}, \\
\langle y'^{2}_{2} \rangle = t'^{2}_{2}V_{m} + \sigma'^{2}_{2}.
\end{eqnarray} 
Estimators $\hat{t}'^{2}_{1}$, $\hat{t}'^{2}_{2}$, $\hat{\sigma}'^{2}_{1}$ and $\hat{\sigma}'^{2}_{2}$ in maximum-likelihood analysis under the normal linear model can be expressed as:
\begin{eqnarray}
\label{array16}
\hat{t}'_{1} &=& \frac{\sum^{m}_{i=1}x_{1i}y'_{1i}}{\sum^{m}_{i=1}x_{1i}^{2}}, \nonumber \\
\hat{t}'_{2} &=& \frac{\sum^{m}_{i=1}x_{2i}y'_{2i}}{\sum^{m}_{i=1}x_{2i}^{2}}. \\
\hat{\sigma}'^{2}_{1} &=& \frac{1}{m}\sum^{m}_{1}(y'_{1i} - \hat{t}'_{1}x_{1i}), \nonumber \\
\hat{\sigma}'^{2}_{2} &=& \frac{1}{m}\sum^{m}_{1}(y'_{2i} - \hat{t}'_{2}x_{2i}).
\end{eqnarray} 
The independent estimators $\hat{t}'_{1}$, $\hat{t}'_{2}$, $\hat{\sigma}'^{2}_{1}$ and $\hat{\sigma}'^{2}_{2}$ follow the distribution below:
\begin{eqnarray}
\label{array17}
\hat{t}'_{1} \sim N(t'_{1}, \frac{\sigma'^{2}_{1}}{\sum^{m}_{i=1}x_{1i}^{2}}), \quad \hat{t}'_{2} &\sim& N(t'_{2}, \frac{\sigma'^{2}_{2}}{\sum^{m}_{i=1}x_{2i}^{2}}). \\
\frac{m\sigma'^{2}_{1}}{\sigma'^{2}_{1}}, \frac{m\sigma'^{2}_{2}}{\sigma'^{2}_{2}} &\sim& \chi^{2} (m - 1).
\end{eqnarray} 
where $t'_{1}$, $t'_{2}$, $\sigma'^{2}_{1}$ and $\sigma'^{2}_{2}$ are the true values of the parameters. The confidence interval of these parameters can be estimated with the except probability $\epsilon_{PE}/2$ due to the limit of $m$ as:
\begin{eqnarray}
\label{array18}
t'_{1} &\in& [t'_{1} - \Delta t'_{1},  t'_{1} + \Delta t'_{1}], \nonumber \\
t'_{2} &\in& [t'_{2} - \Delta t'_{2},  t'_{2} + \Delta t'_{2}], \nonumber \\
\sigma'^{2}_{1} &\in& [\sigma'^{2}_{1} - \Delta \sigma'^{2}_{1}, \sigma'^{2}_{1} + \Delta \sigma'^{2}_{1}], \nonumber \\
\sigma'^{2}_{2} &\in& [\sigma'^{2}_{2} - \Delta \sigma'^{2}_{2}, \sigma'^{2}_{2} + \Delta \sigma'^{2}_{2}].
\end{eqnarray} 
where
\begin{eqnarray}
\label{array19}
\Delta t'_{1} = z_{\epsilon_{PE}/2}\sqrt{\frac{\hat{\sigma}'^{2}_{1}}{mV_{m}}}, \quad \Delta t'_{2} = z_{\epsilon_{PE}/2}\sqrt{\frac{\hat{\sigma}'^{2}_{2}}{mV_{m}}}, \\
\Delta \sigma'^{2}_{1} = z_{\epsilon_{PE}/2}\frac{\sigma'^{2}_{1}}{\sqrt{\frac{m}{2}}}, \quad \Delta \sigma'^{2}_{2} = z_{\epsilon_{PE}/2}\frac{\sigma'^{2}_{2}}{\sqrt{\frac{m}{2}}}.
\end{eqnarray}
where $z_{\epsilon_{PE}/2}$ is around 6.5 when the $\sigma_{PE}$ is generally taken as $10^{-10}$ \cite{leverrier2010finite, jouguet2012analysis}.
Now we can estimate minimum $\eta_{A,x} = \hat{t}'^{2}_{1}$, $\eta_{B,x} = \hat{t}'^{2}_{2}$ and maximum $\epsilon_{A,x} = \frac{\hat{\sigma}'^{2}_{1} - 1}{\hat{t}'^{2}_{1}}$, $\epsilon_{B,x} = \frac{\hat{\sigma}'^{2}_{2} - 1}{\hat{t}'^{2}_{2}}$ using the previous confidence intervals and calculation results. After Charlie has finished the measurements, we can further estimate the parameters in covariance matrix $\gamma_{A_{1}B'_{1}}$ as:
\begin{eqnarray}
\label{array20}
T_{A,x} = \frac{\eta_{A,x}}{2}g^{2}, \quad T_{B,x} = \frac{\eta_{B,x}}{2}g^{2}, \\
\epsilon'_{A,x} =  \epsilon_{A,x} + \frac{2}{\eta_{A,x}} +\frac{\eta_{B,x}}{\eta_{A,x}}(\epsilon_{B,x} - 2).
\end{eqnarray}
where we select $g^{2} = \frac{2V_{m}}{\eta_{B,x}(V_{m}+2)}$. As the parameters are all derived from the above part, now we can analyze the finite-size effects.

\begin{figure}[!h]\center
	\resizebox{12cm}{!}{
		\includegraphics{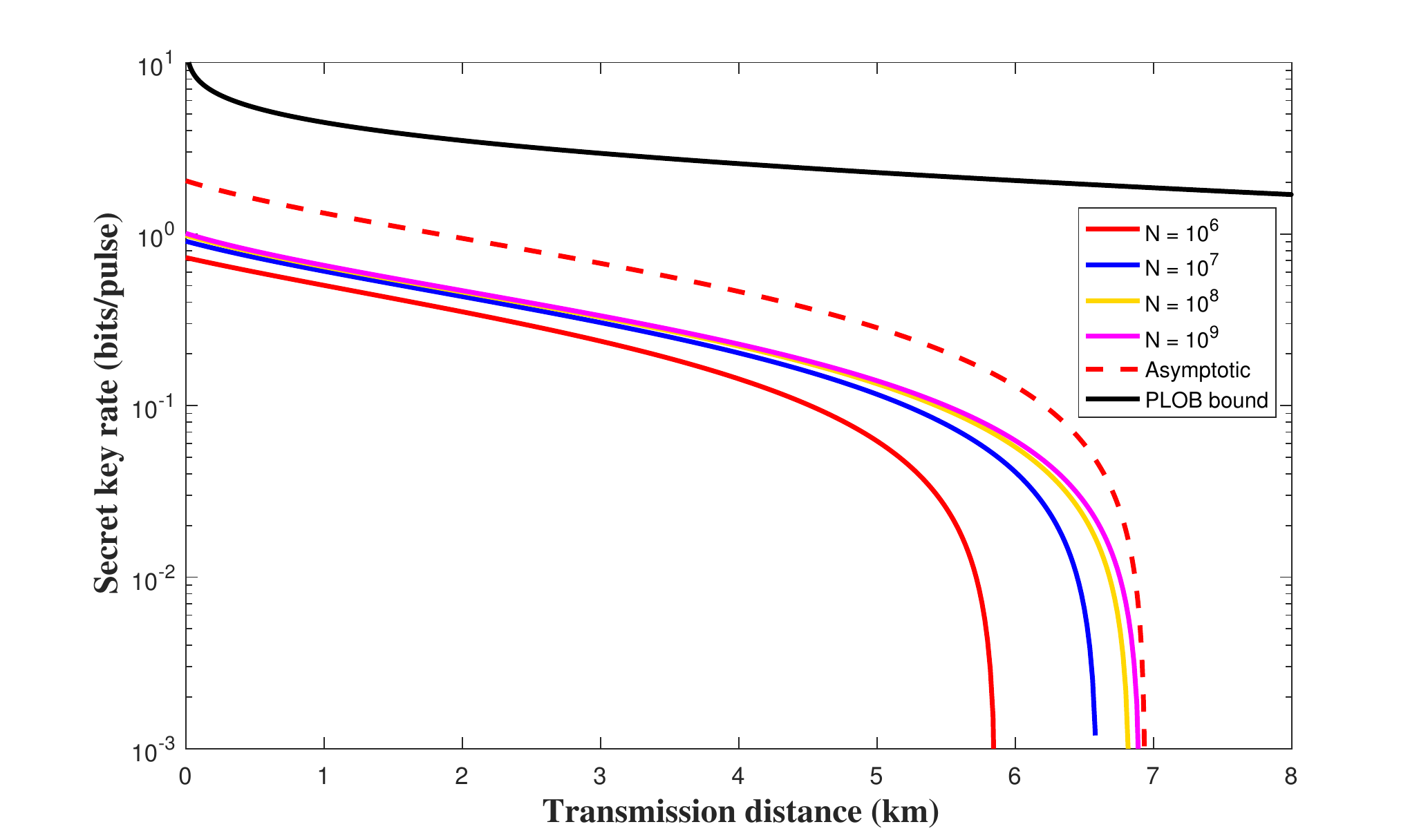}}
	\caption{Secret key rate as a function of transmission distance in symmetric case considering finite-size effects. From left to right, the block length is equal to $10^{6}$, $10^{7}$, $10^{8}$, $10^{9}$ and infinite. The upper solid black line is the PLOB bound. The modulation variance $V_{m}$ is 100, $\beta=0.98$, the excess noise are $\epsilon_{A,x} = \epsilon_{B,x} = 0.002$, the quantum channel loss is 0.2 dB/km.}
	\label{fig:2.8}       % Give a unique label
\end{figure}

\begin{figure}[!h]\center
	\resizebox{12cm}{!}{
		\includegraphics{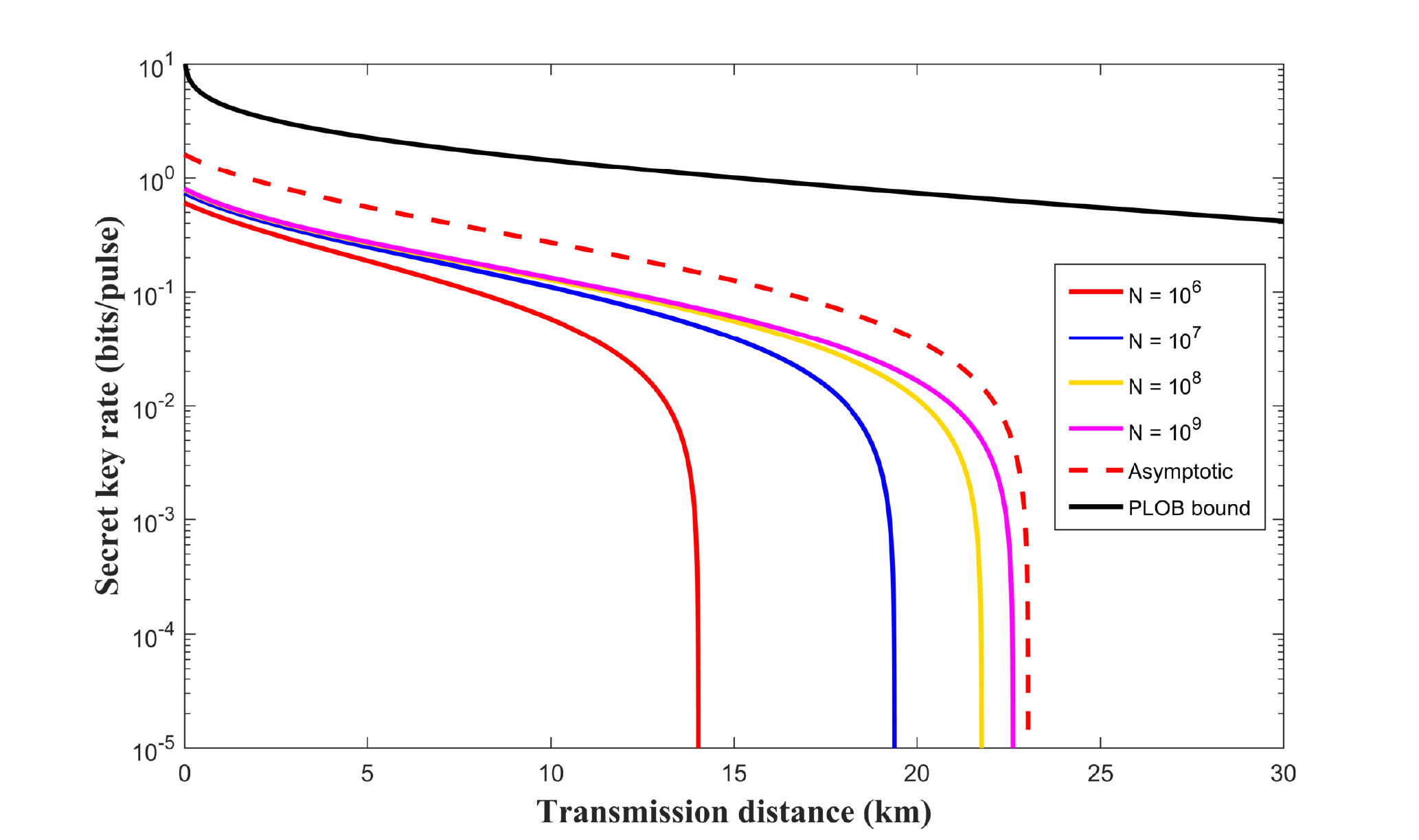}}
	\caption{Secret key rate as a function of transmission distance in asymmetric case considering finite-size effects. Where Charlie is set extremely close to Bob with an overall efficiency 0.98. From left to right, the block length is equal to $10^{6}$, $10^{7}$, $10^{8}$, $10^{9}$ and infinite. The other parameters are the same as Fig. \ref{fig:2.8}. }
	\label{fig:2.9}       % Give a unique label
\end{figure}

Figure. \ref{fig:2.8} and Fig. \ref{fig:2.9} demonstrate the secret key rate as a function of transmission distance in the symmetric case and asymmetric case considering the finite-size effects. The data length $n$ to generate the secret key is half of the total block length. Simulation results show that the finite-size effects will significantly influence the performance of our proposed UD CV-MDI QKD protocol with a rather small amount of data exchanged. As the number of the exchanged data increases, the performance will gradually approach the corresponding asymptotic case. In addition, the scheme is robust against the finite-size effect with the block length larger than $10^{9}$. In both curves, the PLOB bound is plotted and the results are strictly under the PLOB bound region. To better perform UD CV-QKD protocol, it's essential to exchange a sufficient number of block data.

\section{Conclusion}
\label{Con}

In this paper, we have firstly introduced a CV-MDI QKD scheme with unidimensional modulation based on the Gaussian modulation of a single quadrature of the coherent lights, which will greatly reduce the implementation complexity and allow more standard devices, as well we  illustrate the physicality of the other unmodulated quadrature. Moreover, we investigate the finite-size effects under practical conditions to fill the gap between the asymptotic case and the practical case and we found that our protocol is robust to finite-size effects with large block data length (larger than $10^{9}$). Overall, our simulation results under accessible parameters show that compared with the original, symmetric Gaussian modulation protocol, our UD CV-MDI QKD protocol is still comparable to its counterpart with acceptable secret key rate and considerable system simplification. 
 
\begin{acknowledgements}
This work was supported by the National Key Research and Development Program (Grant No. 2016YFA0302600), the National Natural Science Foundation of China (Grants No. 61332019, No. 61671287, No. 61631014), and the National Key Research and Development Program of China (Grant No. 2013CB338002).
\end{acknowledgements}


\bibliographystyle{spmpscinew}      % mathematics and physical sciences
\bibliography{ref}   % name your BibTeX data base

\begin{thebibliography}{10}
\providecommand{\url}[1]{{#1}}
\providecommand{\urlprefix}{URL }
\expandafter\ifx\csname urlstyle\endcsname\relax
  \providecommand{\doi}[1]{DOI~\discretionary{}{}{}#1}\else
  \providecommand{\doi}{DOI~\discretionary{}{}{}\begingroup
  \urlstyle{rm}\Url}\fi

\bibitem{bennett1984update}
Bennett, C.H., Brassard, G.: An update on quantum cryptography.
\newblock In: Workshop on the Theory and Application of Cryptographic
  Techniques, pp. 475--480. Springer (1984)

\bibitem{ekert1991quantum}
Ekert, A.: Quantum cryptography based on bell's theorem.
\newblock Phys. Rev. Lett. \textbf{67}(6), 661--663 (1991)

\bibitem{gisin2002quantum}
Gisin, N., Ribordy, G., Tittel, W., Zbinden, H.: Quantum cryptography.
\newblock Rev. Mod. Phys. \textbf{74}(1), 145 (2002)

\bibitem{scarani2009the}
Scarani, V., Bechmannpasquinucci, H., Cerf, N., Dusek, M., Lutkenhaus, N.,
  Peev, M.: The security of practical quantum key distribution.
\newblock Rev. Mod. Phys \textbf{81}(3), 1301--1350 (2009)

\bibitem{braunstein2005quantum}
Braunstein, S.L., Van~Loock, P.: Quantum information with continuous variables.
\newblock Rev. Mod. Phys. \textbf{77}(2), 513 (2005)

\bibitem{Liao2017Satellite}
Liao, S.K., Cai, W.Q., Liu, W.Y., Zhang, L., Li, Y., Ren, J.G., Yin, J., Shen,
  Q., Cao, Y., Li, Z.P.: Satellite-to-ground quantum key distribution.
\newblock Nature \textbf{549}(7670), 43 (2017)

\bibitem{bang2006quantum}
Bang, J.Y., Berger, M.S.: Quantum mechanics and the generalized uncertainty
  principle.
\newblock Phys. Rev. D \textbf{74}(12), 125012 (2006)

\bibitem{ralph1999continuous}
Ralph, T.C.: Continuous variable quantum cryptography.
\newblock Phys. Rev. A \textbf{61}(1), 010303 (1999)

\bibitem{grosshans2002continuous}
Grosshans, F., Grangier, P.: Continuous variable quantum cryptography using
  coherent states.
\newblock Phys. Rev. Lett. \textbf{88}(5), 057902 (2002)

\bibitem{grosshans2003quantum}
Grosshans, F., Van~Assche, G., Wenger, J., Brouri, R., Cerf, N., Grangier, P.:
  Quantum key distribution using gaussian-modulated coherent states.
\newblock Nature (London) \textbf{421}(6920), 238--241 (2003)

\bibitem{bai2017performance}
Bai, D., Huang, P., Ma, H., Wang, T., Zeng, G.: Performance improvement of
  plug-and-play dual-phase-modulated quantum key distribution by using a
  noiseless amplifier.
\newblock Entropy \textbf{19}(10), 546 (2017)

\bibitem{liu2018integrating}
Liu, W., Huang, P., Peng, J., Fan, J., Zeng, G.: Integrating machine learning
  to achieve an automatic parameter prediction for practical
  continuous-variable quantum key distribution.
\newblock Phys. Rev. A \textbf{97}(2) (2018)

\bibitem{lo2005decoy}
Lo, H.K., Ma, X., Chen, K.: Decoy state quantum key distribution.
\newblock Phys. Rev. lett. \textbf{94}(23), 230504 (2005)

\bibitem{xuan2009a}
Xuan, Q.D., Zhang, Z., Voss, P.L.: A 24 km fiber-based discretely signaled
  continuous variable quantum key distribution system.
\newblock Opt. Express \textbf{17}(26), 24244--24249 (2009)

\bibitem{lo2014secure}
Lo, H.K., Curty, M., Tamaki, K.: Secure quantum key distribution.
\newblock Nat. Photon. \textbf{8}(8), 595 (2014)

\bibitem{gottesman2003secure}
Gottesman, D., Preskill, J.: Secure quantum key distribution using squeezed
  states.
\newblock In: Quantum Information with Continuous Variables, pp. 317--356.
  Springer (2003)

\bibitem{garcia2009continuous}
Garc{\'\i}a-Patr{\'o}n, R., Cerf, N.J.: Continuous-variable quantum key
  distribution protocols over noisy channels.
\newblock Phys. Rev. Lett. \textbf{102}(13), 130501 (2009)

\bibitem{weedbrook2012gaussian}
Weedbrook, C., Pirandola, S., Garciapatron, R., Cerf, N., Ralph, T.C., Shapiro,
  J.H., Lloyd, S.: Gaussian quantum information.
\newblock Rev. Mod. Phys. \textbf{84}(2), 621--669 (2012)

\bibitem{garcia2006unconditional}
Garcia-Patron, R., Cerf, N.J.: Unconditional optimality of gaussian attacks
  against continuous-variable quantum key distribution.
\newblock Phys. Rev. Lett. \textbf{97}(19), 190503 (2006)

\bibitem{navascues2006optimality}
Navascu{\'e}s, M., Grosshans, F., Acin, A.: Optimality of gaussian attacks in
  continuous-variable quantum cryptography.
\newblock Phys. Rev. Lett. \textbf{97}(19), 190502 (2006)

\bibitem{furrer2012continuous}
Furrer, F., Franz, T., Berta, M., Leverrier, A., Scholz, V.B., Tomamichel, M.,
  Werner, R.F.: Continuous variable quantum key distribution: finite-key
  analysis of composable security against coherent attacks.
\newblock Phys. Rev. Lett. \textbf{109}(10), 100502 (2012)

\bibitem{leverrier2013security}
Leverrier, A., Garc{\'\i}a-Patr{\'o}n, R., Renner, R., Cerf, N.J.: Security of
  continuous-variable quantum key distribution against general attacks.
\newblock Phys. Rev. Lett. \textbf{110}(3), 030502 (2013)

\bibitem{leverrier2010finite}
Leverrier, A., Grosshans, F., Grangier, P.: Finite-size analysis of a
  continuous-variable quantum key distribution.
\newblock Phys. Rev. A \textbf{81}(6), 062343 (2010)

\bibitem{jouguet2012analysis}
Jouguet, P., Kunz-Jacques, S., Diamanti, E., Leverrier, A.: Analysis of
  imperfections in practical continuous-variable quantum key distribution.
\newblock Phys. Rev. A \textbf{86}(3), 032309 (2012)

\bibitem{leverrier2015composable}
Leverrier, A.: Composable security proof for continuous-variable quantum key
  distribution with coherent states.
\newblock Phys. Rev. Lett. \textbf{114}(7), 070501 (2015)

\bibitem{lodewyck2007quantum}
Lodewyck, J., Bloch, M., Garc{\'\i}a-Patr{\'o}n, R., Fossier, S., Karpov, E.,
  Diamanti, E., Debuisschert, T., Cerf, N.J., Tualle-Brouri, R., McLaughlin,
  S.W., et~al.: Quantum key distribution over 25 km with an all-fiber
  continuous-variable system.
\newblock Phys. Rev. A \textbf{76}(4), 042305 (2007)

\bibitem{jouguet2013experimental}
Jouguet, P., Kunzjacques, S., Leverrier, A., Grangier, P., Diamanti, E.:
  Experimental demonstration of long-distance continuous-variable quantum key
  distribution.
\newblock Nat. Photon. \textbf{7}(5), 378--381 (2013)

\bibitem{qi2015generating}
Qi, B., Lougovski, P., Pooser, R., Grice, W., Bobrek, M.: Generating the local
  oscillator “locally” in continuous-variable quantum key distribution
  based on coherent detection.
\newblock Phys. Rev. X \textbf{5}(4), 041009 (2015)

\bibitem{wang2018high}
Wang, T., Huang, P., Zhou, Y., Liu, W., Ma, H., Wang, S., Zeng, G.: High key
  rate continuous-variable quantum key distribution with a real local
  oscillator.
\newblock Opt. Express \textbf{26}(3), 2794--2806 (2018)

\bibitem{fossier2009field}
Fossier, S., Diamanti, E., Debuisschert, T., Villing, A., Tualle-Brouri, R.,
  Grangier, P.: Field test of a continuous-variable quantum key distribution
  prototype.
\newblock New J. Phys. \textbf{11}(4), 045023 (2009)

\bibitem{jouguet2012field}
Jouguet, P., Kunz-Jacques, S., Debuisschert, T., Fossier, S., Diamanti, E.,
  All{\'e}aume, R., Tualle-Brouri, R., Grangier, P., Leverrier, A., Pache, P.,
  et~al.: Field test of classical symmetric encryption with continuous
  variables quantum key distribution.
\newblock Opt. Express \textbf{20}(13), 14030--14041 (2012)

\bibitem{huang2016field}
Huang, D., Huang, P., Li, H., Wang, T., Zhou, Y., Zeng, G.: Field demonstration
  of a continuous-variable quantum key distribution network.
\newblock Opt. lett. \textbf{41}(15), 3511--3514 (2016)

\bibitem{huang2016long-distance}
Huang, D., Huang, P., Lin, D., Zeng, G.: Long-distance continuous-variable
  quantum key distribution by controlling excess noise.
\newblock Sci. Rep. \textbf{6}(1), 19201--19201 (2016)

\bibitem{gerhardt2011full}
Gerhardt, I., Liu, Q., Lamas-Linares, A., Skaar, J., Kurtsiefer, C., Makarov,
  V.: Full-field implementation of a perfect eavesdropper on a quantum
  cryptography system.
\newblock Nat. Commun. \textbf{2}, 349 (2011)

\bibitem{Ma2013Local}
Ma, X.C., Sun, S.H., Jiang, M.S., Liang, L.M.: Local oscillator fluctuation
  opens a loophole for eve in practical continuous-variable
  quantum-key-distribution systems.
\newblock Phys. Rev. A \textbf{88}(2), 290--296 (2013)

\bibitem{Jouguet2013Preventing}
Jouguet, P., Kunzjacques, S., Diamanti, E.: Preventing calibration attacks on
  the local oscillator in continuous-variable quantum key distribution.
\newblock Phys. Rev. A \textbf{87}(6), 4996--4996 (2013)

\bibitem{qin2013saturation}
Qin, H., Kumar, R., All{\'e}aume, R.: Saturation attack on continuous-variable
  quantum key distribution system.
\newblock In: Emerging Technologies in Security and Defence; and Quantum
  Security II; and Unmanned Sensor Systems X, vol. 8899, p. 88990N.
  International Society for Optics and Photonics (2013)

\bibitem{QHHDBACVQKD2018}
Qin, H., Kumar, R., Makarov, V., All\'eaume, R.: Homodyne-detector-blinding
  attack in continuous-variable quantum key distribution.
\newblock Phys. Rev. A \textbf{98}, 012312 (2018)

\bibitem{Braunstein2012Side}
Braunstein, S.L., Pirandola, S.: Side-channel-free quantum key distribution.
\newblock Phys. Rev. Lett. \textbf{108}(13), 130502 (2012)

\bibitem{lo2012measurement-device-independent}
Lo, H., Curty, M., Qi, B.: Measurement-device-independent quantum key
  distribution.
\newblock Phys. Rev. Lett. \textbf{108}(13), 130503 (2012)

\bibitem{pirandola2015high-rate}
Pirandola, S., Ottaviani, C., Spedalieri, G., Weedbrook, C., Braunstein, S.L.,
  Lloyd, S., Gehring, T., Jacobsen, C.S., Andersen, U.L.: High-rate
  measurement-device-independent quantum cryptography.
\newblock Nat. Photonics \textbf{9}(6), 397--402 (2015)

\bibitem{Ma2013Gaussian}
Ma, X.C., Sun, S.H., Jiang, M.S., Gui, M., Liang, L.M.: Gaussian-modulated
  coherent-state measurement-device-independent quantum key distribution.
\newblock Phys. Rev. A \textbf{89}(4), 4089--4091 (2013)

\bibitem{li2014continuous-variable}
Li, Z., Zhang, Y., Xu, F., Peng, X., Guo, H.: Continuous-variable
  measurement-device-independent quantum key distribution.
\newblock Phys. Rev. A \textbf{89}(5), 052301 (2014)

\bibitem{ma2018continuous}
Ma, H.X., Huang, P., Bai, D.Y., Wang, S.Y., Bao, W.S., Zeng, G.H.:
  Continuous-variable measurement-device-independent quantum key distribution
  with photon subtraction.
\newblock Phys. Rev. A \textbf{97}(4), 042329 (2018)

\bibitem{zhao2018continuous}
Zhao, Y., Zhang, Y., Xu, B., Yu, S., Guo, H.: Continuous-variable
  measurement-device-independent quantum key distribution with virtual photon
  subtraction.
\newblock Phys. Rev. A \textbf{97}(4), 042328 (2018)

\bibitem{wang2018self}
Wang, Y., Wang, X., Li, J., Huang, D., Zhang, L., Guo, Y.: Self-referenced
  continuous-variable measurement-device-independent quantum key distribution.
\newblock Phys. Lett. A \textbf{382}(17), 1149--1156 (2018)

\bibitem{yin2019phase}
Yin, H.L., Zhu, W., Fu, Y.: Phase self-aligned continuous-variable
  measurement-device-independent quantum key distribution.
\newblock Sci. Rep. \textbf{9}(1), 49 (2019)

\bibitem{ma2019long}
Ma, H.X., Huang, P., Bai, D.Y., Wang, T., Wang, S.Y., Bao, W.S., Zeng, G.H.:
  Long-distance continuous-variable measurement-device-independent quantum key
  distribution with discrete modulation.
\newblock Phys. Rev. A \textbf{99}(2), 022322 (2019)

\bibitem{bai2019passive}
Bai, D., Huang, P., Ma, H., Wang, T., Zeng, G.: Passive state preparation in
  continuous-variable measurement-device-independent quantum key distribution.
\newblock J. Phys. B  (2019)

\bibitem{papanastasiou2017finite}
Papanastasiou, P., Ottaviani, C., Pirandola, S.: Finite-size analysis of
  measurement-device-independent quantum cryptography with continuous
  variables.
\newblock Phys. Rev. A \textbf{96}(4), 042332 (2017)

\bibitem{zhang2017finite}
Zhang, X., Zhang, Y., Zhao, Y., Wang, X., Yu, S., Guo, H.: Finite-size analysis
  of continuous-variable measurement-device-independent quantum key
  distribution.
\newblock Phys. Rev. A \textbf{96}(4), 042334 (2017)

\bibitem{lupo2018continuous}
Lupo, C., Ottaviani, C., Papanastasiou, P., Pirandola, S.: Continuous-variable
  measurement-device-independent quantum key distribution: Composable security
  against coherent attacks.
\newblock Phys. Rev. A \textbf{97}(5), 052327 (2018)

\bibitem{usenko2015unidimensional}
Usenko, V.C., Grosshans, F.: Unidimensional continuous-variable quantum key
  distribution.
\newblock Phys. Rev. A \textbf{92}(6), 062337 (2015)

\bibitem{wang2017finite}
Wang, P., Wang, X., Li, J., Li, Y.: Finite-size analysis of unidimensional
  continuous-variable quantum key distribution under realistic conditions.
\newblock Opt. Express \textbf{25}(23), 27995--28009 (2017)

\bibitem{liao2018composable}
Liao, Q., Guo, Y., Xie, C., Huang, D., Huang, P., Zeng, G.: Composable security
  of unidimensional continuous-variable quantum key distribution.
\newblock Quantum Inf. Process. \textbf{17}(5), 113 (2018)

\bibitem{wang2018security}
Wang, P., Wang, X., Li, Y.: Security analysis of unidimensional
  continuous-variable quantum key distribution using uncertainty relations.
\newblock Entropy \textbf{20}(3), 157 (2018)

\bibitem{wang2017experimental}
Wang, X., Liu, W., Wang, P., Li, Y.: Experimental study on all-fiber-based
  unidimensional continuous-variable quantum key distribution.
\newblock Phys.l Rev. A \textbf{95}(6), 062330 (2017)

\bibitem{fossier2009improvement}
Fossier, S., Diamanti, E., Debuisschert, T., Tuallebrouri, R., Grangier, P.:
  Improvement of continuous-variable quantum key distribution systems by using
  optical preamplifiers.
\newblock J. Phys. B \textbf{42}(11), 114014 (2009)

\bibitem{Pirandola2017Fundamental}
Pirandola, S., Laurenza, R., Ottaviani, C., Banchi, L.: Fundamental limits of
  repeaterless quantum communications.
\newblock Nat. Commun. \textbf{8}, 15043 (2017)

\bibitem{Ruppert2014Long}
Ruppert, L., Usenko, V.C., Filip, R.: Long-distance continuous-variable quantum
  key distribution with efficient channel estimation.
\newblock Phys. Rev. A \textbf{90}(6), 062310 (2014)

\end{thebibliography}

%\begin{thebibliography}{}
%%
%% and use \bibitem to create references. Consult the Instructions
%% for authors for reference list style.
%\end{thebibliography}

\end{document}